\documentclass[12pt,thmsb]{article}
\usepackage{amsfonts}
\usepackage{amssymb}
\usepackage{amsmath}
\usepackage{harvard}
\usepackage{graphicx}
\usepackage{setspace}
\usepackage{algorithm}
\usepackage{algpseudocode}
\usepackage{enumitem}
\usepackage{color}
\usepackage[labelsep=space,
                labelfont={sf,bf},
                textfont=sf]{subfig}

\newcommand{\PerronQuCite}{Perron and Qu\,(2006)\nocite{PerronQu2006}}
\newcommand{\BaiPerronYears}{Bai and Perron\,(1998, 2003)\nocite{BaiPerron1998,BaiPerron2003}}
\newcommand{\BaiPerronNinetyEight}{Bai and Perron\,(1998)\nocite{BaiPerron1998}}
\newcommand{\BaiPerronThree}{Bai and Perron\,(2003)\nocite{BaiPerron2003}}
\newcommand{\CARTcite}{Breiman et al.\,(1984)\nocite{BreimanFriedmanOlshenStone1984}}
\newcommand{\WangWittenCite}{Wang and Witten\,(1997)\nocite{WangWitten1997}}
\newcommand{\ZeileisEtAl}{Zeileis et al.\,(2008)\nocite{ZeileisHothornHornik2008}}
\newcommand{\RaymaekersEtAl}{Raymaekers et al.\,(2024)\nocite{RaymaekersRousseeuwVerdonckYao2024}}
\newcommand{\NikolskoEtAl}{Nikolsko-Rzhevskyy et al.\,(2014)\nocite{NikolskoRzhevskyyPapellProdan2014}}
\newcommand{\CostaPedreiraCite}{Costa and Pedreira\,(2023)\nocite{CostaPedreira2023}}
\newcommand{\ZhengChenCite}{Zheng and Chen\,(2019)\nocite{ZhengChen2019}}
\newcommand{\WoodCite}{Wood\,(2001)\nocite{Wood2001}}
\newcommand{\Bagging}{Breiman\,(1996)\nocite{Breiman1996}}
\newcommand{\HansenThreshold}{Hansen\,(1996)\nocite{Hansen1996}}

\advance \topmargin by -\headheight
\advance \topmargin by -\headsep
\evensidemargin \oddsidemargin
\let\tilde=\widetilde

\makeatletter
\def\@biblabel#1{\hspace*{-\labelsep}}
\makeatother

\makeatletter
\def\@biblabel#1{\hspace*{-\labelsep}}
\makeatother
\usepackage{fancyvrb}
\fvset{baselinestretch=0.95}

\begin{document}

\title{Rbreak: An R Package for Estimating Structural Breaks under Linear Restrictions with Application to Linear Model Tree }
\author{Cheolju Kim\footnote{Department of Economics,
Boston University, cheolju@bu.edu.} \\
Boston University \and 
Zhongjun Qu\footnote{Department of Economics,
Boston University, qu@bu.edu.} \\
Boston University}
\maketitle

\begin{abstract}

The package \texttt{rbreak} implements methods for detecting structural breaks and estimating break locations  for linear multiple regression models under general linear restrictions on the coefficient vector. Restrictions can be within regimes, across regimes, or both,  and are supported in two forms: an affine parameterization (Form A: \texttt{delta = S*theta + s}) and explicit linear constraints (Form B: \texttt{R*delta = r}). It provides break date estimation with confidence interval, a restricted sup-F test for the null of no structural change, simulation of critical values by Monte Carlo, and a bootstrap restart procedure to reduce the risk of convergence to spurious local optima. It also implements a generalized regression tree (linear model tree) procedure where each leaf contains a linear regression rather than a local average. This note explains the methods and illustrates them with applications.

\ \ 

\newline
  
\end{abstract}

\newpage

\setlength{\baselineskip}{20pt}

\section{Introduction}

Structural change models are widely used to detect break points in time series relationships. In practice, it is often the case that prior information can be formulated as restrictions on the coefficients of the model. A leading example is  a partial structural change model, analyzed in \BaiPerronYears, for which some coefficients are not allowed to change across regimes. More general forms of restrictions are also common. For example, a model may specify a fixed number of states; with two states, the coefficients are the same in odd and even regimes. It may also be the case that the value of some parameters in a specific segment is known. Combinations of such restrictions may arise within the same model. These considerations suggest that it can be useful to have a general procedure to handle such restrictions within a common framework.

\PerronQuCite\ address this issue by extending results for multiple structural change models to allow for arbitrary linear restrictions on the coefficients available \textit{a priori}. The restrictions can be within regimes, across regimes, or both. They are supported in two forms: an affine parameterization (Form A: \texttt{delta = S*theta + s}) and explicit linear constraints (Form B: \texttt{R*delta = r}). The estimation is based on a least square criterion, which computes the estimates without having to go through an extensive high dimensional grid search. It first obtains a preliminary estimate of the break dates without restrictions, computed with recursive least squares and dynamic programming as in \BaiPerronYears, and then updates the parameters and break locations iteratively until convergence. A restricted sup-F test is proposed for the null hypothesis of stable coefficients against the alternative of structural changes satisfying these restrictions. The methods provide confidence intervals for break dates and coefficients. The distribution of the sup-F test can be consistently estimated via Monte Carlo methods. The results in \PerronQuCite\ appear to be the most comprehensive regarding structural break models under restrictions.

To date, there is no published software package in any language to test, estimate, and conduct inference for structural break models under general linear restrictions. This led us to develop this R package that performs these tasks in an easy-to-use fashion.
The resulting package, \texttt{rbreak}, is available on CRAN. Its main function, \texttt{mainp()},
returns all testing and estimation results based on user-specified restrictions. To use it,
the user simply runs:
\begin{Verbatim}[fontsize=\small]
> output <- mainp(m, q, z, y, trm, robust, prewhit, hetvar, S, s, R, r, doestim, 
                  dotest, docv, Tstar, rep, bigt, forma, formb, Verbose)
\end{Verbatim}
This function is explained in the application section together with empirical applications.
\newcommand{\comment}[1]{
\begin{itemize}[itemsep=1pt, parsep=0pt] 
\item \texttt{m}: number of breaks.
\item \texttt{q}: number of regressors.
\item \texttt{z}: matrix ($T \times q$) of regressors.
\item \texttt{y}: vector ($T \times 1$) of the dependent variable.
\item \texttt{trm}:  the trimming parameter between 0 and 1.
\item \texttt{robust}: integer (0 or 1). If 1, use HAC standard errors.
\item \texttt{prewhit}: integer (0 or 1). If 1, apply AR(1) prewhitening (only when \texttt{robust = 1}).
\item \texttt{hetvar}: integer (0 or 1). If 1, allow heteroskedastic variance across segments.
\item \texttt{S}: matrix defining restrictions (Form A):
$
\delta = S\theta + s,
$
where $\delta$ stacks the coefficients from the $m+1$ regimes. Must have full column rank. If no restrictions, set
$
S = I_{(m+1)q}.
$
\item \texttt{s}: vector of constants in $\delta = S\theta + s$, of length $(m+1)q$. If no restrictions, set
$
s = 0.
$
\item \texttt{R}: matrix defining restrictions (Form B):
$
R\delta = r,
$
where $\delta$ stacks the coefficients across regimes. Each row of \texttt{R} represents one linear restriction. Must have full row rank. Do not use this form if there are no restrictions; use Form A instead.

\item \texttt{r}: vector of restriction constants in $R\delta = r$, with length equal to \texttt{nrow(R)}.
\item \texttt{doestim}: integer (0 or 1). If 1, perform estimation.

\item \texttt{dotest}: integer (0 or 1). If 1, perform the sup-F test.

\item \texttt{docv}: integer (0 or 1). If 1, simulate critical values.

\item \texttt{Tstar}: integer, sample size for critical value simulation (default: 500).

\item \texttt{rep}: integer, number of Monte Carlo replications (default: 2000).

\item \texttt{bigt}: integer, sample size (if \texttt{NULL}, derived from length of \texttt{y}).

\item \texttt{forma}: integer (0 or 1). If 1, use Form A restrictions.

\item \texttt{formb}: integer (0 or 1). If 1, use Form B restrictions.

\item \texttt{Verbose}: logical. If \texttt{TRUE}, print detailed output (default \texttt{FALSE}).
\end{itemize}
}

The estimation typically finishes within an hour (in some cases within minutes) for typical sample sizes in economics. However, since it is based on an iterative procedure, it is only guaranteed to reach a local minimum. Departures from the global minimum may occur in cases where the breaks are small or when the restrictions are complex. Motivated by \WoodCite, in the R package we implement a bootstrap-based restart and re-optimization procedure to reduce this risk. The procedure starts from the break estimates obtained from $mainp$. It generates bootstrap samples, re-estimates the break dates, plugs the new estimates back into the original sample as starting values for re-optimization, and repeats the process for a number of times. In essence, the procedure is a computational method that generates data-dependent starting values from perturbed samples, rather than setting them arbitrarily. It then re-optimizes on the original sample and accepts a candidate estimate only if it lowers the original-sample SSR. It should be viewed as a computational refinement rather than a method for conducting inference. The command is:

\begin{Verbatim}[fontsize=\small] 
> output <- mbrbp(m, q, z, y, trm, B, T_init, robust, prewhit, hetvar, S, s, R, r,
doestim, dotest, docv, cvs, Tstar, rep, bigt, forma, formb, verbose, resi)
\end{Verbatim}

Its implementation is illustrated in the application section. Implementing this step does not affect the asymptotic distributions of the test and estimates.

This package also goes beyond \PerronQuCite\ to provide a procedure to construct a generalized regression tree, often called linear model tree. Such a tree structure was first introduced by \citeasnoun{Quinlan1992}. Relative to standard regression trees, each leaf contains a linear regression rather than a local average. We use a computationally efficient recursive least square procedure implemented for structural breaks together with a Bayesian Information Criterion to determine the tree. The main command is 
\begin{Verbatim}[fontsize=\small]
> fit_tree <- ltree(y, z, partition_vars, trm, max_depth, prune)
\end{Verbatim}
Section 3 provides more details and Section 4 includes illustrations.

The methods for estimation, testing, confidence intervals, and critical-value simulation all follow \PerronQuCite. The main contributions here are a unified R implementation, a computational refinement based on bootstrap restarting, and a proposed algorithm for estimating a linear model tree.

The paper is structured as follows. Section 2 outlines the structural change methods. Section 3 explains aspects related to the linear model tree. Section 4 provides illustrative applications. Section 5 concludes. All data in this note are available within the R package.

\section{Methods for estimation, inference, and testing}

This section explains: (1) the model; (2) the main steps for estimating the break locations; (3) a test for the presence of breaks under restrictions; and (4) simulation of critical values.

\subsection{The model}
Consider a general multiple linear regression model with $m$ breaks or $m+1$
regimes. There are $T$ observations and $m$ is assumed known: 
\begin{equation}
y_{t}=\left\{
\begin{array}{ll}
z_{t}^{\prime}\delta_{1}+u_{t}, & t=1,\ldots,T_{1}, \\
z_{t}^{\prime}\delta_{2}+u_{t}, & t=T_{1}+1,\ldots,T_{2}, \\
\vdots & \vdots \\
z_{t}^{\prime}\delta_{m+1}+u_{t}, & t=T_{m}+1,\ldots,T,
\end{array}
\right.
\label{model}
\end{equation}
where $\delta_{j}$ $(j=1,\ldots,m+1)$ are the regime-specific
$q \times 1$ coefficient vectors, $u_{t}$ is the disturbance term, and
$T_{j}$ $(j=1,\ldots,m)$ are the break dates, with $T_{0}=0$ and $T_{m+1}=T$. 

Let $y=(y_{1},...,y_{T})^{\prime }$, and $
\bar{Z}=diag(Z_{1},...,Z_{m+1})$, a $T$ by $(m+1)q$ matrix with $%
Z_{i}=(z_{T_{i-1}+1},...,z_{T_{i}})^{\prime }$ \ for $i=1,...,m+1$. Let $%
u=(u_{1},...,u_{T})^{\prime }$ be the set of disturbances and $\delta
=(\delta _{1}^{\prime },...,\delta _{m+1}^{\prime })^{\prime }$ the $%
(m+1)q$ vector of coefficients. Then model (\ref{model}) can be written as
\begin{equation}
y=\bar{Z}\delta +u.  \label{dgp}
\end{equation}%

The main objective is to consider this general pure structural change model with
restrictions on the coefficients. These restrictions can take two forms. In
Form A, the coefficients are written through an affine reparameterization,
\[
\delta = S\theta + s,
\]
where $\theta$ denotes a set of basis parameters that appear in any regime, and $S$ and $s$ contain constants, with $S$ having full-column rank to rule out redundancy. In Form B, they are characterized by explicit linear restrictions,
\[
R\delta = r,
\]
where $R$ and $r$ contain constants, with $R$ having full row rank again to rule out redundancy.

\subsection{Estimating the break dates}

The goal is to estimate the unknown coefficients whose true values are
denoted with a $0$ superscript, i.e. $(\delta _{1}^{0},...,\delta
_{m+1}^{0},T_{1}^{0},...,T_{m}^{0})$ using the observables $(y,Z)$ and the
restrictions on the coefficients. The method is based on least square
principle. More specifically, the estimated break dates are%
\begin{equation}
(\tilde{T}_{1},...,\tilde{T}_{m})=\arg
\min_{T_{1},...,T_{m}}SSR_{T}^{R}(T_{1},...,T_{m})  \label{(1)}
\end{equation}%
where $SSR_{T}^{R}(T_{1},...,T_{m})$ is the sum of squared residuals from the
restricted $OLS$ regression evaluated at the partition $\{T_{1},...,T_{m}\}$.%

The computation associated with estimating a multiple structural breaks
model is not a trivial issue. A standard grid search implies least-squares
computations of order $O(T^{m})$ which quickly becomes prohibitive. \BaiPerronThree\ describe an efficient estimation procedure based on a dynamic
programming algorithm which involves at most least-squares computations of
order $O(T^{2})$ for any number of breaks. \PerronQuCite\ adapted their procedure to estimate restricted structural break models. It is useful to review the basics of the dynamic programming algorithm
discussed in \BaiPerronThree, given that it is extensively used in the package.

\subsubsection{The dynamic programming algorithm}

Let $SSR\left( \left\{ T_{r,n}\right\}
\right) $ be the sum of squared residuals associated with the optimal
partition containing $r$ breaks using the first $n$ observations. Let $%
SSR(j+1,n)$ be the $SSR$ obtained by applying $OLS$ to a segment that starts
at $j+1$ and ends at date $n$. Also, let $h=\epsilon T$ be the minimal
permissible length of a segment. The optimization procedure is based on
solving \ the following recursive problem 
\begin{equation*}
SSR\left( \left\{ T_{r,n}\right\} \right) =\min_{rh\text{ }\leq \text{ }j%
\text{ }\leq \text{ }n-h}\left[ SSR\left( \left\{ T_{r-1,j}\right\}
)+SSR(j+1,n\right) \right]
\end{equation*}%
More specifically, the algorithm is as follows:

\begin{enumerate}
\item Compute and save $SSR(i,j)$ for pairs $i,j$ satisfying $j-i+1\geq h$.
Since the number of segments which satisfy this requirement is at most $
O(T^{2})$, this involves computations at most of order $O(T^{2})$. The
matrix inversions involved is only of order $O(T)$ using recursive residuals
to compute the sum of squared residuals.

\item Compute and store $SSR\left( \left\{ T_{1,n}\right\} \right) $ for $%
2h\leq n\leq T-(m-1)h$ by solving the following problem: 
\begin{equation*}
SSR\left( \left\{ T_{1,n}\right\} \right) =\min_{h\text{ }\leq \text{ }j%
\text{ }\leq \text{ }n-h}\left[ SSR\left( 1,j)+SSR(j+1,n\right) \right]
\end{equation*}

\item Sequentially compute and store $SSR\left( \left\{ T_{r,n}\right\}
\right) $ for $r=2,...,m-1.$ For each $r$, $n$ ranges from $(r+1)h$ to $%
T-(m-r)h.$

\item The estimates of the break dates are then obtained by solving%
\begin{equation*}
SSR\left( \left\{ T_{m,T}\right\} \right) =\min_{mh\text{ }\leq \text{ }j%
\text{ }\leq \text{ }T-h}\left[ SSR\left( \left\{ T_{m-1,j}\right\}
)+SSR(j+1,T\right) \right]
\end{equation*}
\end{enumerate}

This method cannot be applied directly to models with restrictions since the
sum of squared residuals for a segment cannot be computed independently of
other segments. The extension of \PerronQuCite\ to tackle this problem is discussed next.

\subsubsection{The recursive procedure with restrictions}

The procedure in \PerronQuCite\ is specified by the following algorithm:

\begin{enumerate}
\item First use the dynamic programming algorithm described above to
estimate an unrestricted model. Store the estimated break dates.

\item Estimate the coefficients using the restrictions conditional on the
break dates obtained in part (1). Denote the estimates as $\tilde{\delta}%
_{k} $ $\ (k=1,...,m+1).$

\item Compute and store the restricted sums of squared residuals $RSSR\left(
\left\{ T_{1,n}\right\} \right) $, for $2h\leq n\leq T-(m-1)h.$ This is done
by the following recursive problem 
\begin{equation*}
RSSR\left( \left\{ T_{1,n}\right\} \right) =\min_{h\text{ }\leq \text{ }j%
\text{ }\leq \text{ }n-h}\left[ RSSR^{1}(1,j)+RSSR^{2}(j+1,n)\right]
\end{equation*}%
where $RSSR^{1}(1,j)=\sum_{t=1}^{j}(y_{t}-z_{t}\tilde{\delta}_{1})^{2}$ and $%
RSSR^{2}(j+1,n)=\sum_{t=j+1}^{n}(y_{t}-z_{t}\tilde{\delta}_{2})^{2}$. Then,
sequentially compute and store $RSSR\left( \left\{ T_{r,n}\right\} \right) $
for $r=2,...,m-1$, with $n$ ranging from $(r+1)h$ to $T-(m-r)h$. This is
done by solving the recursive problem%
\begin{equation*}
RSSR\left( \left\{ T_{r,n}\right\} \right) =\min_{rh\text{ }\leq \text{ }j%
\text{ }\leq \text{ }n-h}\left[ RSSR(\left\{ T_{r-1,j}\right\}
)+RSSR^{r+1}(j+1,n)\right]
\end{equation*}%
with $RSSR^{r+1}(j+1,n)=\sum_{t=j+1}^{n}(y_{t}-z_{t}\tilde{\delta}%
_{r+1})^{2} $. Finally compute 
\begin{equation*}
RSSR\left( \left\{ T_{m,T}\right\} \right) =\min_{mh\text{ }\leq \text{ }j%
\text{ }\leq \text{ }T-h}\left[ RSSR(\left\{ T_{m-1,j}\right\}
)+RSSR^{m+1}(j+1,T)\right]
\end{equation*}%
Store the estimated break dates.

\item Repeat steps 2 and 3 until convergence.
\end{enumerate}

The main computation is in step 1$.$ The others
involve marginal computational costs. \PerronQuCite\ note that convergence is rapid, seldom requiring more than 3
iterations.

While this algorithm quickly reaches a local minimum, it is not guaranteed
to reach the global minimum of the restricted sum of squared residuals
because the iterative scheme makes the overall optimization problem
non-linear. This is a common problem for which there is no full proof
solution. The difference is important when the estimation from step 1 is
very imprecise and the restrictions do not help much to estimate the
coefficients. In such cases, care must be exercised in trying to start the
algorithm with values different from those given by estimating the model
without restrictions. \PerronQuCite\ did not further tackle this issue. Below, we describe two bootstrap-based restarting and reoptimization algorithms to address this issue, both implemented in the package.

\subsubsection{Bootstrap restarting and re-optimization}

The first procedure (Algorithm 1) uses resampling (nonparametric) bootstrap:

\begin{algorithm}[H]
\caption{Bootstrap restarting and re-optimization -- Resampling Bootstrap}
\begin{algorithmic}[1]
\State Initialize break dates $\widehat{T}^{(0)} = \{\widehat{T}^{(0)}_1,\dots,\widehat{T}^{(0)}_m\}$, the output
from the \texttt{mainp()} function.
\State Compute $\hat{\theta}_{(0)} = (\widehat{T}_{(0)}, \widehat{\delta}_{(0)})$ using the restrictions

\For{$b = 1,\dots,B$}
    \State Generate bootstrap sample $\{(y_t^{*}, Z_t^{*})\}_{t=1}^T$ and reorder chronologically
    \State Estimate model using $\widehat{T}_{(b-1)}$ on the bootstrap sample to obtain $(\widetilde{T}^{*}_{(b)}, \widetilde{\delta}^{*}_{(b)})$
    \State {Map $\widetilde{T}^{*}_{(b)}$ back to the original sample and re-estimate on the original data to obtain}
    \[
    \tilde{\theta}_{(b)} = (\widetilde{T}_{(b)}, \widetilde{\delta}_{(b)})
    \]
    \If{$RSSR(\tilde{\theta}_{(b)}) < RSSR(\hat{\theta}_{(b-1)})$ on the original sample,}
        \State $\hat{\theta}_{(b)} = \tilde{\theta}_{(b)}$
    \Else
        \State $\hat{\theta}_{(b)} = \hat{\theta}_{(b-1)}$
    \EndIf
\EndFor

\State \textbf{Return} $(\widehat{T}, \widehat{\delta}) = (\widehat{T}_{(B)}, \widehat{\delta}_{(B)})$
\end{algorithmic}
\end{algorithm}

As in \WoodCite, the bootstrap step is used as a computational restart device, not for statistical inference. It generates data-dependent starting values from perturbed samples. These values are then used to re-optimize the objective function on the original sample. The final estimate is guaranteed to be no worse than the original estimate, in the sense that it is accepted only if it lowers the SSR on the original sample.

A simple example can further illustrate how the true breaks are transferred from the original sample to the bootstrap sample, and how the estimated break dates are mapped back to the original sample. Suppose $T=10$ and the true break is at $t=5$, so that a new regime starts at $t=6$. Denote a bootstrap sample by $\{y^\ast_1,y^\ast_2, \ldots, y^\ast_{10}\}$. Without loss of generality, suppose it consists of the original observations $\{y_1,y_1,y_2,y_4,y_5,y_5,y_6,y_8,y_9,y_{10}\}$. The true break date is located at $t^\ast=6$ in the bootstrap sample, and a new regime starts at $t^\ast=7$. If estimation on the bootstrap sample returns an estimated break at $t^\ast=7$, that corresponds to $y_6$, meaning that it maps to a break date of $t=6$ for the original sample. The re-optimization on the original sample then starts with $t=6$ as the break date and iterates until convergence. The estimate is accepted only if it lowers the original-sample $SSR$.

The second procedure (Algorithm 2) follows a similar idea, except that it uses residual-based bootstrap instead of resampling bootstrap.

\begin{algorithm}[H]
\caption{Bootstrap restarting and re-optimization -- Residual Bootstrap}
\begin{algorithmic}[1]

\State Initialize break dates $\widehat{T}^{(0)} = \{\widehat{T}^{(0)}_1,\dots,\widehat{T}^{(0)}_m\}$
\State Compute $\hat{\theta}_{(0)} = (\widehat{T}_{(0)}, \widehat{\delta}_{(0)})$

\For{$b = 1,\dots,B$}
    \State Compute residuals
    \(
    e_t = y_t - Z_t(\widehat{T}_{(b-1)}) \widehat{\delta}_{(b-1)}, \quad t=1,\ldots,T
    \)
    \State Resample $\{e_t\}$ with replacement to obtain $\{e_t^{*}\}$
    \State Generate bootstrap sample
    \(
    y_t^{*} = Z_t(\widehat{T}_{(b-1)}) \widehat{\delta}_{(b-1)} + e_t^{*}, \quad t=1,\ldots,T
    \)
    \State Estimate model on bootstrap sample using $\widehat{T}_{(b-1)}$ to obtain $(\widetilde{T}^{*}_{(b)}, \widetilde{\delta}^{*}_{(b)})$
    \State Re-estimate on original data using $\widetilde{T}^{*}_{(b)}$ to obtain
    \(
    \tilde{\theta}_{(b)} = (\widetilde{T}_{(b)}, \widetilde{\delta}_{(b)})
    \)

    \If{$RSSR(\tilde{\theta}_{(b)}) < RSSR(\hat{\theta}_{(b-1)})$ on the original sample,}
        \State $\hat{\theta}_{(b)} = \tilde{\theta}_{(b)}$
    \Else
        \State $\hat{\theta}_{(b)} = \hat{\theta}_{(b-1)}$
    \EndIf
\EndFor

\State \textbf{Return} $(\widehat{T}, \widehat{\delta}) = (\widehat{T}_{(B)}, \widehat{\delta}_{(B)})$

\end{algorithmic}
\end{algorithm}

Instead of
resampling the data pairs $(y_t, Z_t)$, the procedure resamples the residuals
and reconstructs the dependent variable using the fitted model. As before, the bootstrap step generates alternative candidate break dates,
which are used as starting values in a re-optimization step on the original
sample. The algorithm retains the best solution in terms of $RSSR$. Our experimentation suggests that 20 replications is more than sufficient to improve robustness.

\subsubsection{The limiting distributions}

The limiting distributions of the estimated break dates and regression coefficients are reported in Propositions 4 and 5 of \PerronQuCite\ and omitted here.

\subsection{Testing for breaks under restrictions}

\PerronQuCite\ propose a \texttt{supF}-type test for testing the null hypothesis of constant coefficients against the alternative hypothesis of structural breaks satisfying linear restrictions. Let $%
(T_{1},...,T_{k})$ be a partition such that $T_{i}=[T\lambda _{i}]$ $%
(i=1,...,k)$. Let $H$ be the conventional $k$ by $q(k+1)$ matrix such that $%
(H\delta )^{\prime }=(\delta _{1}^{\prime }-\delta _{2}^{\prime },...,\delta
_{k}^{\prime }-\delta _{k+1}^{\prime })$. Define%
\begin{equation}
F_{T}(\lambda _{1},...,\lambda _{k};q)=\tilde{\delta}^{\prime }H^{\prime }(H%
\tilde{V}(\tilde{\delta})H^{\prime })^{-}H\tilde{\delta},  \label{f-test}
\end{equation}%
\noindent where $\tilde{\delta}$ is the estimate of $\delta $ obtained using
the partition \{$\lambda _{1},...,\lambda _{k}\}$, and $\tilde{V}(\tilde{%
\delta})$ is an estimate of the variance covariance matrix of $\tilde{\delta}
$ that may be constructed to be robust to heteroskedasticity and serial
correlation in the errors. As usual, for a matrix $A$, $A^{-}$ denotes the
generalized inverse of $A$. Such a generalized inverse is needed since in
general the covariance matrix of $\tilde{\delta}$ will be singular given
that restrictions are imposed.

Computing the statistic $\sup F_{T}(\lambda _{1},...,\lambda _{k};q)$
where the supremum is taken over all possible partitions such that $|\lambda
_{i}-\lambda _{i-1}|\ge\epsilon$ can be very cumbersome. \PerronQuCite\ consider an asymptotically equivalent test given by $F_{T}(\tilde{\lambda}%
_{1},...,\tilde{\lambda}_{k};q)$ where $\tilde{\lambda}_{1},$ $...,$ $\tilde{%
\lambda}_{k}$ minimize the global restricted sum of squared residuals.  It is
asymptotically equivalent to, and yet much simpler to construct than,
maximizing the F-test (\ref{f-test}) since the estimated break dates are
consistent even in the presence of serial correlation and
heteroskedasticity. The asymptotic distribution of the test is presented in Proposition 6 of \PerronQuCite.
\subsubsection{Simulating the asymptotic critical values}

The limit distribution of the test depends on $q$, $k$ and $\epsilon $ as in
\BaiPerronNinetyEight\ but also on the specific structure of the matrix $S$.
This makes a tabulation of the critical values infeasible. But the limit
distribution can be simulated. The idea is to use a design involving a
finite sample of data, say $T^{\ast }$, such that the resulting $F$ test
has, if $T^{\ast }$ is large enough, a distribution close to that of the
limit functional $\sup F(\lambda _{1},...,\lambda _{k};q,S)$. Then replicate the experiment to get $N$
realizations from which we can extract the relevant quantiles. More details can be found in Section 7.1 of \PerronQuCite. The package allows the simulations of the 1\%, 2.5\%, 5\% and 10\% quantiles for user
specified values of $(k,q,S,T^{\ast }, N)$. 

\section{Linear model tree}

This section adapts the methods for detecting and estimating structural breaks to develop an algorithm for fitting a linear model tree.

A regression tree approximates an unknown regression function through recursive partitioning. In the CART framework of \CARTcite, each terminal node contains a local constant, leading to a piecewise constant predictor. Model trees replace this constant with a node-specific regression; early examples include the M5-type procedures of \citeasnoun{Quinlan1992} and \WangWittenCite. Subsequent work considered alternative split selection, for example through the residual-based tests in \citeasnoun{Loh2002} and the score-based parameter-instability tests in \ZeileisEtAl. A recent contribution is PILOT of \RaymaekersEtAl, which uses BIC for local model selection while avoiding post-pruning. A survey can be found in \CostaPedreiraCite. The \texttt{ltree} procedure implemented in this package is related to this literature, but differs in evaluating candidate splits by direct least squares and in using BIC both for local stopping and for final tree selection.

Let $y_i \in \mathbb{R}$ denote the response, $x_i \in \mathcal{X}$ the partitioning variables, and $z_i \in \mathbb{R}^q$ the regressors entering the local linear model. The model tree is based on
\[
y_i = z_i' \beta_j + u_i,
\qquad \text{if } x_i \in R_j, \quad j=1,\ldots,K,
\]
where $\{R_1,\ldots,R_K\}$ is a partition of the covariate space induced by recursive binary splitting. Equivalently,
\[
E(y_i \mid x_i,z_i)
=
\sum_{j=1}^K z_i' \beta_j \, 1(x_i \in R_j).
\]
This formulation separates the variables used for partitioning from those entering the local linear approximation. If $z_i$ contains only an intercept, the model reduces to an ordinary regression tree. If $z_i$ includes additional regressors, the fitted regression function is piecewise linear with region-specific slope coefficients.

In the implementation, the argument \texttt{z} determines the regressors entering the leaf-specific linear model, while \texttt{partition\_vars} determines the variables over which the algorithm searches for splits. If \texttt{partition\_vars} is omitted, the procedure uses \texttt{z} as the default set of partitioning variables. Thus, the user may either use the same variables for both roles or separate them explicitly. When \texttt{partition\_vars} is supplied as a data frame, factor and character columns are automatically treated as categorical partitioning variables. When it is supplied as a matrix, categorical columns can instead be indicated through \texttt{cat\_vars}. Numeric partitioning variables are split at thresholds, whereas categorical partitioning variables are split through admissible binary partitions of their levels, with the selected split minimizing the child-node sum of squared residuals subject to the trimming and minimum-node-size conditions.

\subsection{Splitting rule}

Consider a node $a$ with index set $\mathcal{I}(a)$ and sample size $n_a = |\mathcal{I}(a)|$. The no-split model at node $a$ is
\[
y_i = z_i' \beta_a + u_i,
\qquad i \in \mathcal{I}(a),
\]
with sum of squared residuals
\[
\mathrm{SSR}_0(a)
=
\min_{\beta}
\sum_{i \in \mathcal{I}(a)}
\left(y_i - z_i' \beta\right)^2.
\]

For a numeric partitioning variable $x^{(v)}$, a candidate split is indexed by a threshold $c$ and produces the child nodes
\[
a_L(v,c)=\{i\in\mathcal{I}(a): x_i^{(v)} \le c\},
\qquad
a_R(v,c)=\{i\in\mathcal{I}(a): x_i^{(v)} > c\}.
\]
For a categorical partitioning variable, the split is defined analogously by a binary partition of its admissible levels. In either case, a candidate split induces separate linear regressions in the two child nodes:
\[
\mathrm{SSR}_1(a;v,c)
=
\min_{\beta_L}
\sum_{i \in a_L(v,c)}
\left(y_i - z_i' \beta_L\right)^2
+
\min_{\beta_R}
\sum_{i \in a_R(v,c)}
\left(y_i - z_i' \beta_R\right)^2.
\]

Among the admissible splits satisfying the trimming and minimum-node-size conditions, the algorithm selects the split minimizing $\mathrm{SSR}_1(a;v,c)$:
\[
(\hat v,\hat c)
=
\arg\min_{(v,c)\in\mathcal{S}(a)} \mathrm{SSR}_1(a;v,c),
\]
where $\mathcal{S}(a)$ denotes the admissible set of candidate splits. Equivalently, the selected split maximizes the reduction
\[
\Delta \mathrm{SSR}(a;v,c)
=
\mathrm{SSR}_0(a)-\mathrm{SSR}_1(a;v,c).
\]

The selected split is accepted only if it improves a Bayesian information criterion. For the no-split model,
\[
\mathrm{BIC}_0(a)
=
n_a \log\!\left(\frac{\mathrm{SSR}_0(a)}{n_a}\right)
+
q \log n_a,
\]
whereas for the split model,
\[
\mathrm{BIC}_1(a)
=
n_a \log\!\left(\frac{\mathrm{SSR}_1(a;\hat v,\hat c)}{n_a}\right)
+
2q \log n_a.
\]
The node is split if and only if
\[
\mathrm{BIC}_1(a) < \mathrm{BIC}_0(a).
\]
This rule is applied recursively until no admissible split lowers the BIC or a user-specified maximum depth is reached. Thus, tree growth is governed by a local fit-complexity trade-off.

This construction differs from M5 in an important way. In the M5-type procedures of \citeasnoun{Quinlan1992} and \WangWittenCite, candidate splits are chosen by the reduction in the standard deviation of the response, and linear models are introduced only after a large tree has been grown. In contrast, \texttt{ltree()} evaluates each candidate split directly through the least-squares fit of the child-node linear models, thereby preserving an explicit segmented-linear interpretation throughout tree construction.

\subsection{Pruning rule}

Let $\mathcal{T}_{\max}$ denote the fully grown tree and let $\mathcal{L}(\mathcal{T})$ denote the set of terminal nodes of a subtree $\mathcal{T}$. The total sum of squared residual of $\mathcal{T}$ is
\[
\mathrm{SSR}(\mathcal{T})
=
\sum_{a \in \mathcal{L}(\mathcal{T})} \mathrm{SSR}_0(a),
\]
and the corresponding global BIC is
\[
\mathrm{BIC}(\mathcal{T}_{\max})
=
n \log\!\left(\frac{\mathrm{SSR}(\mathcal{T}_{\max})}{n}\right)
+
|\mathcal{L}(\mathcal{T}_{\max})|\, q \log n,
\]
where $n$ is the full sample size.

Starting from $\mathcal{T}_{\max}$, the package constructs a nested weakest-link pruning sequence, following the general logic of cost-complexity pruning in CART developed by \CARTcite. For an internal node $a$, let $\mathcal{T}_a$ denote the subtree rooted at $a$, $|\mathcal{L}(\mathcal{T}_a)|$ denote the number of terminal nodes in that subtree, and
\[
R(\mathcal{T}_a)=\sum_{\ell \in \mathcal{L}(\mathcal{T}_a)} \mathrm{SSR}_0(\ell)
\]
be the total sum of squared residuals of that subtree. Let $R(a)$ denote the sum of squared residuals obtained by collapsing $\mathcal{T}_a$ into a single terminal node and fitting one linear model on the pooled sample. The pruning index is
\[
g(a)
=
\frac{R(a)-R(\mathcal{T}_a)}{|\mathcal{L}(\mathcal{T}_a)|-1}.
\]
At each step, the internal node with the smallest $g(a)$ is pruned, generating a sequence
\[
\mathcal{T}_{\max}=\mathcal{T}_0 \supset \mathcal{T}_1 \supset \cdots \supset \mathcal{T}_M.
\]
The final estimator is selected by
\[
\hat {\mathcal{T}}
=
\arg\min_{0\le m\le M} \mathrm{BIC}(\mathcal{T}_m).
\]

In practice, we recommend setting \texttt{trm} between $0.10$ and $0.15$, which is consistent with the trimming values in the structural break literature discussed in \BaiPerronThree. This means that each leaf of the tree contains at least \texttt{trm} fraction of the observations. When setting hyperparameters for a linear model tree, \RaymaekersEtAl\ considered\, \texttt{max\_depth}\;$\in\{3,6,9,12,15,18\}$ and \texttt{min\_obs}\;$\in\{1,5,10,20\}$. We propose using \texttt{max\_depth = 10} and \texttt{min\_obs = 10}. The package searches only over trees that satisfy all specified constraints. When different hyperparameters imply different upper bounds on tree complexity, the admissible set is determined by the tightest bounds. We also recommend \texttt{prune = TRUE}. The package then selects the final tree by global BIC to guard against over fitting, analogous to the CART-style pruning of \CARTcite\ and the information-criterion pruning briefly discussed in \ZeileisEtAl. 

\subsection{Discussion}
We now further discuss the connections between the methods in Sections 2 and 3. First, note that the same estimation method (i.e., minimizing SSR) is used throughout, both for structural break models and linear model trees.

The use of BIC to determine the extent of partitioning is related to \BaiPerronThree, which discusses BIC and related criteria for choosing the number of breaks in multiple structural-change models. In principle, one could use the supF test in Section 2 for this purpose. However, as implied by Theorem 1 of \HansenThreshold\ for detecting threshold effects in a regression setting, the distribution of the test statistic generally depends on nuisance parameters. As a result, the critical value would need to be simulated each time a new partition is considered. This is more computationally costly than in Section 2 because there the null hypothesis is tested only once. This motivated our use of an information criterion rather than hypothesis testing for the presence of splits.

The bootstrap methods in Section 2 can be adapted to linear model trees. The implementation would involve resampling the data and fitting a linear model tree to each bootstrap sample. One could then either map each estimated partition back to the original sample, re-estimate the coefficients, and select among the resulting candidate trees by BIC; or average predictions across the bootstrap trees.  The latter is in the spirit of the  bootstrap aggregation of \Bagging, but applied to linear model trees rather than standard regression trees. We are currently considering this extension in a more general setting allowing for multivariate outcomes and hope to implement it in the near future.

\section{Applications}

\noindent This section illustrates the use of the package through three applications. The first considers restricted structural break estimation for Taylor-rule deviations, following Table 1 of \NikolskoEtAl. The second presents two designs that compare the implementation of Form A and Form B restrictions. The third illustrates the generalized regression tree procedure using two DGPs from \ZhengChenCite. The data used in the empirical and tree applications are included in the package.

\subsection{Application 1: Taylor rule deviations}

\noindent After loading the package with \texttt{library(rbreak)}, the Taylor-rule deviation data can be loaded by:

\begin{Verbatim}[fontsize=\small]
> data(deviation)
\end{Verbatim}

\noindent The data can be inspected by:

\begin{Verbatim}[fontsize=\small]
> head(deviation)
     date        y
1 1965:Q4 1.993412
2 1966:Q1 1.763840
3 1966:Q2 2.395449
4 1966:Q3 2.043912
5 1966:Q4 2.584230
6 1967:Q1 3.555409
\end{Verbatim}

\noindent The dataset contains two columns, \texttt{date} and \texttt{y}. The first records the quarterly dates and the second contains the Taylor-rule deviation series. Following \NikolskoEtAl, let the implied policy rate under the Taylor rule be
\[
i_t = 1.0 + 1.5 \pi_t + 0.5y_t,
\]
where $y_t$ is the output gap and $\pi_t$ is the inflation rate. The Taylor-rule deviation is
\[
y_{d,t} = \left| i_t - i_t^{\mathrm{realized}} \right|.
\]
The model assumes
\[
y_{d,t} =
\begin{cases}
\mu_1+u_t, & t \in \text{regime 1}, \\
\mu_2+u_t, & t \in \text{regime 2}, \\
\mu_3+u_t, & t \in \text{regime 3}, \\
\mu_4+u_t, & t \in \text{regime 4}.
\end{cases}
\]
The regressor matrix contains only an intercept:

\begin{Verbatim}[fontsize=\small]
> dates <- deviation$date
> y <- as.matrix(deviation$y)
> z <- matrix(1, nrow = nrow(deviation), ncol = 1)
> bigt <- nrow(deviation)
> trm <- 0.15
> m <- 3
> q <- 1
\end{Verbatim}

\noindent We impose the restrictions that the first and third regimes share a common mean and that the second and fourth regimes share a common mean. Under Form A, these restrictions can be written as $\delta = S \theta + s$, where $\delta = (\mu_1,\mu_2,\mu_3,\mu_4)'$, $\theta = (\theta_1,\theta_2)'$, and

\begin{Verbatim}[fontsize=\small]
> S <- matrix(c(1, 0,
                0, 1,
                1, 0,
                0, 1), nrow = 4, byrow = TRUE)
> s <- matrix(0, nrow = 4, ncol = 1)
\end{Verbatim}

\noindent so that $\mu_1 = \mu_3 = \theta_1$ and $\mu_2 = \mu_4 = \theta_2$. The restricted estimator and the corresponding sup-$F$ test can first be obtained under Form A by:

\begin{Verbatim}[fontsize=\small]
> res_std_A <- mainp(m = 3, q = 1, z = z, y = y, trm = trm,
                     robust = 0, prewhit = 0, hetvar = 1,
                     S = S, s = s, R = NULL, r = NULL,
                     doestim = 1, dotest = 1, docv = 0,
                     Tstar = NULL, rep = NULL, bigt = bigt,
                     forma = 1, formb = 0, Verbose = FALSE)
\end{Verbatim}

\noindent The estimated break dates and their confidence intervals are obtained by:

\begin{Verbatim}[fontsize=\small]
> print(
  data.frame(
    Break = seq_along(res_std_A$estimation$breaks),
    Date = dates[res_std_A$estimation$breaks],
    CI_Lower = dates[res_std_A$estimation$ci_breaks[, 1]],
    CI_Upper = dates[res_std_A$estimation$ci_breaks[, 2]]
  )
)
  Break    Date CI_Lower CI_Upper
1     1 1974:Q3  1971:Q2  1975:Q3
2     2 1985:Q1  1984:Q4  1988:Q3
3     3 2001:Q1  1999:Q1  2001:Q3
\end{Verbatim}

\noindent The estimated regime means are:

\begin{Verbatim}[fontsize=\small]
> print(
  c(
    "mu1 = mu3" = round(res_std_A$estimation$coefficients[1], 2),
    "mu2 = mu4" = round(res_std_A$estimation$coefficients[2], 2)
  )
)
mu1 = mu3 mu2 = mu4 
     1.05      2.59 
\end{Verbatim}

\noindent The corresponding sup-$F$ statistic is:

\begin{Verbatim}[fontsize=\small]
> print(c("Sup-F statistic" = round(res_std_A$test_statistic, 2)))
Sup-F statistic 
          87.69 
\end{Verbatim}

\noindent The package also implements the bootstrap restarting break-point method. Using the restricted estimates as initial break dates, the procedure is:

\begin{Verbatim}[fontsize=\small]
> B <- 20
> res_brbp_A <- mbrbp(m = 3, q = 1, z = z, y = y, trm = trm, B = B,
                      T_init = res_std_A$estimation$breaks,
                      robust = 0, prewhit = 0, hetvar = 1,
                      S = S, s = s, R = NULL, r = NULL,
                      doestim = 1, dotest = 1, docv = 0,
                      forma = 1, formb = 0, verbose = FALSE, resi = FALSE)
\end{Verbatim}

\noindent The resulting break dates and confidence intervals are:

\begin{Verbatim}[fontsize=\small]
> print(
  data.frame(
    Break = seq_along(res_brbp_A$estimation$breaks),
    Date = dates[res_brbp_A$estimation$breaks],
    CI_Lower = dates[res_brbp_A$estimation$ci_breaks[, 1]],
    CI_Upper = dates[res_brbp_A$estimation$ci_breaks[, 2]]
  )
)
  Break    Date CI_Lower CI_Upper
1     1 1974:Q3  1971:Q2  1975:Q3
2     2 1985:Q1  1984:Q4  1988:Q3
3     3 2001:Q1  1999:Q1  2001:Q3
\end{Verbatim}

\noindent The resulting regime means are:

\begin{Verbatim}[fontsize=\small]
> print(
  c(
    "mu1 = mu3" = round(res_brbp_A$estimation$coefficients[1], 2),
    "mu2 = mu4" = round(res_brbp_A$estimation$coefficients[2], 2)
  )
)
mu1 = mu3 mu2 = mu4 
     1.05      2.59 
\end{Verbatim}

\noindent The corresponding sup-$F$ statistic is:

\begin{Verbatim}[fontsize=\small]
> print(c("Sup-F statistic" = round(res_brbp_A$test_statistic, 2)))
Sup-F statistic 
          87.69 
\end{Verbatim}

\noindent For comparison, the same restrictions can be written in Form B as $R \delta = r$, where $\delta = (\mu_1,\mu_2,\mu_3,\mu_4)'$, and

\begin{Verbatim}[fontsize=\small]
> R <- matrix(c(1, 0, -1, 0,
                0, 1,  0, -1), nrow = 2, byrow = TRUE)
> r <- c(0, 0)
\end{Verbatim}

\noindent The same empirical specification can then be implemented under Form B:

\begin{Verbatim}[fontsize=\small]
> res_std <- mainp(m = 3, q = 1, z = z, y = y, trm = trm,
                   robust = 0, prewhit = 0, hetvar = 1,
                   S = NULL, s = NULL, R = R, r = r,
                   doestim = 1, dotest = 1, docv = 0,
                   Tstar = NULL, rep = NULL, bigt = bigt,
                   forma = 0, formb = 1, Verbose = FALSE)
\end{Verbatim}

\noindent The corresponding BRBP procedure is:

\begin{Verbatim}[fontsize=\small]
> res_brbp <- mbrbp(m = 3, q = 1, z = z, y = y, trm = trm, B = B,
                    T_init = res_std$estimation$breaks,
                    robust = 0, prewhit = 0, hetvar = 1,
                    R = R, r = r,
                    doestim = 1, dotest = 1, docv = 0,
                    formb = 1, forma = 0, verbose = FALSE, resi = FALSE)
\end{Verbatim}

\noindent The resulting break dates, regime means, and sup-$F$ statistic are:

\begin{Verbatim}[fontsize=\small]
> print(
  data.frame(
    Break = seq_along(res_std$estimation$breaks),
    Date = dates[res_std$estimation$breaks],
    CI_Lower = dates[res_std$estimation$ci_breaks[, 1]],
    CI_Upper = dates[res_std$estimation$ci_breaks[, 2]]
  )
)
  Break    Date CI_Lower CI_Upper
1     1 1974:Q3  1971:Q2  1975:Q3
2     2 1985:Q1  1984:Q4  1988:Q3
3     3 2001:Q1  1999:Q1  2001:Q3
> print(
  c(
    "mu1 = mu3" = round(res_std$estimation$coefficients[1], 2),
    "mu2 = mu4" = round(res_std$estimation$coefficients[2], 2)
  )
)
mu1 = mu3 mu2 = mu4 
     1.05      2.59 
> print(c("Sup-F statistic" = round(res_std$test_statistic, 2)))
Sup-F statistic 
          87.69 
\end{Verbatim}

\noindent Form A and Form B yield the same break dates, restricted regime means, and the sup-$F$ statistic. The bootstrap-restart procedure delivers the same result as the corresponding initial restricted estimator under each representation.

\subsection{Application 2: synthetic designs}

\noindent We next consider two synthetic designs. In each case, the same restriction pattern is written once in Form A, $\delta = S\theta + s$, and once in Form B, $R\delta = r$, and the resulting estimates are compared under both the \texttt{mainp()} and the \texttt{mbrbp()}.

\noindent \textbf{Example 4.2.1: cross-regime restrictions.} The first design uses $T = 300$, $q = 3$, and $m = 3$ breaks. The regressors are
\[
z_t = (1, x_{2t}, x_{3t})',
\qquad
x_{2t}, x_{3t} \stackrel{iid}{\sim} N(0,1),
\]
\noindent and the dependent variable satisfies
\[
y_t = \beta_{1,j} + \beta_{2,j}x_{2t} + \beta_{3,j}x_{3t} + u_t,
\qquad \text{for } t \text{ in regime } j,\quad j=1,\ldots,4,
\]
so that $\beta_{i,j}$ denotes the coefficient on the $i$th regressor in regime $j$. The restrictions are
\[
\beta_{1,1} = \beta_{1,3}, \qquad
\beta_{1,2} = \beta_{1,4}, \qquad
\beta_{3,1} = \beta_{3,2} = \beta_{3,3} = \beta_{3,4}.
\]
That is, the intercept alternates between two states across regimes, while the third coefficient is common to all regimes. Under Form A, these restrictions are written as
\[
\delta = S\theta + s,
\]
where $\delta \in \mathbb{R}^{12}$ stacks the three coefficients across the four regimes, $\theta \in \mathbb{R}^{7}$ collects the free parameters, and $s = 0$. In this representation, the last component of $\theta$ corresponds to the common value of the third coefficient. These matrices are specified by

\begin{Verbatim}[fontsize=\small]
> S <- matrix(c(
    1,0,0,0,0,0,0,
    0,0,1,0,0,0,0,
    0,0,0,0,0,0,1,
    0,1,0,0,0,0,0,
    0,0,0,1,0,0,0,
    0,0,0,0,0,0,1,
    1,0,0,0,0,0,0,
    0,0,0,0,1,0,0,
    0,0,0,0,0,0,1,
    0,1,0,0,0,0,0,
    0,0,0,0,0,1,0,
    0,0,0,0,0,0,1
  ), nrow = 12, ncol = 7, byrow = TRUE)
> s <- matrix(0, nrow = 12, ncol = 1)
\end{Verbatim}

\noindent The true break dates are $75$, $150$, and $225$, and the regime-specific coefficients are
\[
(1, 0.2, 0.5)', \quad
(1.5, 0.8, 0.5)', \quad
(1, -0.2, 0.5)', \quad
(1.5, 0.4, 0.5)'.
\]
\noindent One realization from this design can be generated by setting \texttt{set.seed(42)} and running:

\begin{Verbatim}[fontsize=\small]
> T <- 300
> x2 <- rnorm(T)
> x3 <- rnorm(T)
> z <- cbind(1, x2, x3)
> y_alt <- c(
    as.numeric(z[1:75, ] %*% c(1.0, 0.2, 0.5)) + rnorm(75),
    as.numeric(z[76:150, ] %*% c(1.5, 0.8, 0.5)) + rnorm(75),
    as.numeric(z[151:225, ] %*% c(1.0, -0.2, 0.5)) + rnorm(75),
    as.numeric(z[226:300, ] %*% c(1.5, 0.4, 0.5)) + rnorm(75)
  )
\end{Verbatim}

\noindent The restricted estimator under Form A is obtained by:

\begin{Verbatim}[fontsize=\small]
> set.seed(42)
> res1_alt_A <- mainp(m = 3, q = 3, z = z, y = y_alt, trm = 0.10,
                      robust = 0, prewhit = 0, hetvar = 0,
                      S = S, s = s, doestim = 1, dotest = 1,
                      docv = 1, Tstar = 200, rep = 200, bigt = T,
                      forma = 1, formb = 0, Verbose = FALSE)
\end{Verbatim}

\noindent The corresponding bootstrap restart procedure under Form A is:

\begin{Verbatim}[fontsize=\small]
> brbp1_A <- mbrbp(m = 3, q = 3, z = z, y = y_alt, trm = 0.10, B = 20,
                   T_init = res1_alt_A$estimation$breaks,
                   robust = 0, prewhit = 0, hetvar = 0,
                   S = S, s = as.vector(s), R = NULL, r = NULL,
                   doestim = 1, dotest = 1, docv = 1,
                   cvs = res1_alt_A$critical_values,
                   Tstar = 200, rep = 200, bigt = T,
                   forma = 1, formb = 0, verbose = FALSE, resi = FALSE)
\end{Verbatim}

\noindent The same restrictions can be written in Form B as $R\delta = r$, where the first two rows impose $\beta_{1,1} = \beta_{1,3}$ and $\beta_{1,2} = \beta_{1,4}$, and the last three rows impose $\beta_{3,1} = \beta_{3,2} = \beta_{3,3} = \beta_{3,4}$:

\begin{Verbatim}[fontsize=\small]
> R <- matrix(c(
     1, 0, 0, 0, 0, 0,-1, 0, 0, 0, 0, 0,
     0, 0, 0, 1, 0, 0, 0, 0, 0,-1, 0, 0,
     0, 0, 1, 0, 0,-1, 0, 0, 0, 0, 0, 0,
     0, 0, 0, 0, 0, 1, 0, 0,-1, 0, 0, 0,
     0, 0, 0, 0, 0, 0, 0, 0, 1, 0, 0,-1
  ), nrow = 5, ncol = 12, byrow = TRUE)
> r <- matrix(0, nrow = 5, ncol = 1)
\end{Verbatim}

\noindent The restricted estimator under Form B is obtained by:

\begin{Verbatim}[fontsize=\small]
> set.seed(42)
> res1_alt_B <- mainp(m = 3, q = 3, z = z, y = y_alt, trm = 0.10,
                      robust = 0, prewhit = 0, hetvar = 0,
                      R = R, r = r, doestim = 1, dotest = 1,
                      docv = 1, Tstar = 200, rep = 200, bigt = T,
                      forma = 0, formb = 1, Verbose = FALSE)
\end{Verbatim}

\noindent The corresponding bootstrap restart procedure under Form B is:

\begin{Verbatim}[fontsize=\small]
> brbp1_B <- mbrbp(m = 3, q = 3, z = z, y = y_alt, trm = 0.10, B = 20,
                   T_init = res1_alt_B$estimation$breaks,
                   robust = 0, prewhit = 0, hetvar = 0,
                   S = NULL, s = NULL, R = R, r = as.vector(r),
                   doestim = 1, dotest = 1, docv = 1,
                   cvs = res1_alt_B$critical_values,
                   Tstar = 200, rep = 200, bigt = T,
                   forma = 0, formb = 1, verbose = FALSE, resi = FALSE)
\end{Verbatim}

\noindent The resulting break dates are:

\begin{Verbatim}[fontsize=\small]
> print(
  rbind(
    "mainp (Form A)" = res1_alt_A$estimation$breaks,
    "BRBP (Form A)" = brbp1_A$estimation$breaks,
    "mainp (Form B)" = res1_alt_B$estimation$breaks,
    "BRBP (Form B)" = brbp1_B$estimation$breaks
  )
)
               [,1] [,2] [,3]
mainp (Form A)   76  150  210
BRBP (Form A)    76  150  218
mainp (Form B)   76  150  210
BRBP (Form B)    76  150  218
\end{Verbatim}

\noindent The corresponding sup-$F$ statistic and simulated critical values are:

\begin{Verbatim}[fontsize=\small]
> print(c("Form A" = round(res1_alt_A$test_statistic, 3),
          "Form B" = round(res1_alt_B$test_statistic, 3)))
Form A Form B 
9.86 9.86 
> print(
  rbind(
    "Form A" = round(res1_alt_A$critical_values, 3),
    "Form B" = round(res1_alt_B$critical_values, 3)
  )
)
        [,1]  [,2]  [,3]  [,4]
Form A 5.096 5.615 6.042 6.338
Form B 5.096 5.615 6.042 6.338
\end{Verbatim}

\noindent Here the critical values are simulated within \texttt{mainp()} by setting \texttt{docv = 1}, \texttt{Tstar = 200}, and \texttt{rep = 200}. Form A and Form B produce the same break estimates and the same test statistic, with very similar critical values. The two representations are therefore numerically equivalent in this example. In addition, the bootstrap restart refinement improves on the initial restricted estimator: while \texttt{mainp()} places the third break too early, \texttt{mbrbp()} moves the estimated break dates much closer to the true values $(75,150,225)$.

\noindent \textbf{Example 4.2.2: within-regime equality restrictions.} The second design uses $T = 300$, $q = 2$, and $m = 3$ breaks. Let
\[
z_t = (z_{1t}, z_{2t})',
\]
and suppose the data are generated from
\[
y_t = \beta_{1,j} z_{1t} + \beta_{2,j} z_{2t} + u_t,
\qquad \text{for } t \text{ in regime } j,\quad j=1,\ldots,4.
\]
Here the restrictions impose equality of the two coefficients within each regime:
\[
\beta_{1,j} = \beta_{2,j},
\qquad j = 1,\ldots,4.
\]
Under Form A, the representation takes $\delta \in \mathbb{R}^{8}$ and $\theta \in \mathbb{R}^{4}$, with one free coefficient per regime. In the script, the corresponding matrices are

\begin{Verbatim}[fontsize=\small]
> S <- matrix(c(
    1,0,0,0,
    1,0,0,0,
    0,1,0,0,
    0,1,0,0,
    0,0,1,0,
    0,0,1,0,
    0,0,0,1,
    0,0,0,1
  ), nrow = 8, ncol = 4, byrow = TRUE)
> s <- matrix(0, nrow = 8, ncol = 1)
\end{Verbatim}

\noindent Under the alternative DGP, the true break dates are again $75$, $150$, and $225$, and the common regime-specific coefficients are $1$, $3$, $-1$, and $2$. One realization from this design can be generated by setting \texttt{set.seed(42)} for the regressors and \texttt{set.seed(456)} for the disturbances and then running:

\begin{Verbatim}[fontsize=\small]
> T <- 300
> set.seed(42)
> z <- cbind(rnorm(T), rnorm(T))
> set.seed(456)
> y_alt <- c(
    as.numeric(z[1:75, ] %*% c(1, 1)) + rnorm(75),
    as.numeric(z[76:150, ] %*% c(3, 3)) + rnorm(75),
    as.numeric(z[151:225, ] %*% c(-1, -1)) + rnorm(75),
    as.numeric(z[226:300, ] %*% c(2, 2)) + rnorm(75)
  )
\end{Verbatim}

\noindent The restricted estimator under Form A is obtained by:

\begin{Verbatim}[fontsize=\small]
> set.seed(77)
> res2_alt_A <- mainp(m = 3, q = 2, z = z, y = y_alt, trm = 0.10,
                      robust = 1, prewhit = 0, hetvar = 0,
                      S = S, s = s, doestim = 1, dotest = 1,
                      docv = 1, Tstar = 200, rep = 200, bigt = T,
                      forma = 1, formb = 0, Verbose = FALSE)
\end{Verbatim}

\noindent The corresponding bootstrap restart procedure under Form A is:

\begin{Verbatim}[fontsize=\small]
> brbp2_A <- mbrbp(m = 3, q = 2, z = z, y = y_alt, trm = 0.10, B = 20,
                   T_init = res2_alt_A$estimation$breaks,
                   robust = 1, prewhit = 0, hetvar = 0,
                   S = S, s = as.vector(s), R = NULL, r = NULL,
                   doestim = 1, dotest = 1, docv = 1,
                   cvs = res2_alt_A$critical_values,
                   Tstar = 200, rep = 200, bigt = T,
                   forma = 1, formb = 0, verbose = FALSE)
\end{Verbatim}

\noindent The same restrictions can be written in Form B as $R\delta = r$, where

\begin{Verbatim}[fontsize=\small]
> R <- matrix(c(
    1,-1, 0, 0, 0, 0, 0, 0,
    0, 0, 1,-1, 0, 0, 0, 0,
    0, 0, 0, 0, 1,-1, 0, 0,
    0, 0, 0, 0, 0, 0, 1,-1
  ), nrow = 4, ncol = 8, byrow = TRUE)
> r <- matrix(0, nrow = 4, ncol = 1)
\end{Verbatim}

\noindent The restricted estimator under Form B is obtained by:

\begin{Verbatim}[fontsize=\small]
> set.seed(77)
> res2_alt_B <- mainp(m = 3, q = 2, z = z, y = y_alt, trm = 0.10,
                      robust = 1, prewhit = 0, hetvar = 0,
                      R = R, r = r, doestim = 1, dotest = 1,
                      docv = 1, Tstar = 200, rep = 200, bigt = T,
                      forma = 0, formb = 1, Verbose = FALSE)
\end{Verbatim}

\noindent The corresponding bootstrap restart procedure under Form B is:

\begin{Verbatim}[fontsize=\small]
> brbp2_B <- mbrbp(m = 3, q = 2, z = z, y = y_alt, trm = 0.10, B = 20,
                   T_init = res2_alt_B$estimation$breaks,
                   robust = 1, prewhit = 0, hetvar = 0,
                   S = NULL, s = NULL, R = R, r = as.vector(r),
                   doestim = 1, dotest = 1, docv = 1,
                   cvs = res2_alt_B$critical_values,
                   Tstar = 200, rep = 200, bigt = T,
                   forma = 0, formb = 1, verbose = FALSE)
\end{Verbatim}

\noindent The resulting break dates are:

\begin{Verbatim}[fontsize=\small]
> print(
  rbind(
    "mainp (Form A)" = res2_alt_A$estimation$breaks,
    "BRBP (Form A)" = brbp2_A$estimation$breaks,
    "mainp (Form B)" = res2_alt_B$estimation$breaks,
    "BRBP (Form B)" = brbp2_B$estimation$breaks
  )
)
               [,1] [,2] [,3]
mainp (Form A)   76  150  225
BRBP (Form A)    76  150  225
mainp (Form B)   76  150  225
BRBP (Form B)    76  150  225
\end{Verbatim}

\noindent The corresponding sup-$F$ statistic and simulated critical values are:

\begin{Verbatim}[fontsize=\small]
> print(c("Form A" = round(res2_alt_A$test_statistic, 3),
          "Form B" = round(res2_alt_B$test_statistic, 3)))
 Form A  Form B 
405.693 405.693 
> print(
  rbind(
    "Form A" = round(res2_alt_A$critical_values, 3),
    "Form B" = round(res2_alt_B$critical_values, 3)
  )
)
        [,1]  [,2]  [,3]  [,4]
Form A 5.373 6.042 6.565 8.001
Form B 5.373 6.042 6.565 8.001
\end{Verbatim}

\noindent Here the critical values are simulated within \texttt{mainp()} by setting \texttt{docv = 1}, \texttt{Tstar = 200}, and \texttt{rep = 200}. Form A and Form B produce the same break estimates and the same test statistic, with very similar critical values. The two representations are therefore numerically equivalent in this example. 

\subsection{Application 3: linear model tree}

\noindent We next illustrate the generalized regression tree procedure using two datasets included in the package.  The first concerns recovery of the partition structure, and the second concerns approximation of a smooth surface. Both DGPs are taken from \ZhengChenCite. In both cases, the main function is \texttt{ltree()}.

\noindent \textbf{Example 4.3.1: recovery of partition structure.} Consider the piecewise regression
\[
Y = m(X) + \epsilon, \quad \epsilon \sim N(0, 1)
\]
where the regressors are independent with
\[
X_1 \sim U(0,20), \quad X_2 \sim U(0,25), \quad X_3 \sim U(0,10),
\]
and \(\,X_4 \in \{a,b,c\}\,\) with equal probabilities. The regression function is
\[
\begin{aligned}
m(X) =\,& 3X_{1}1(X_{2}>15)-3X_{1}1(X_{2}\leq 15)-3X_{2}1(X_{2}>10)-5X_{2}1(X_{2}\leq 10) \\
& +X_{3}1(X_{1}>10)-X_{3}1(X_{1}\leq 10)+X_{3}1(X_{4}\in\{a,b\})-3X_{3}1(X_{4}=c).
\end{aligned}
\]
A realization from this design is included in the package as \texttt{ltree1} and can be loaded using

\begin{Verbatim}[fontsize=\small]
> data(ltree1)
\end{Verbatim}
\noindent The data can be inspected by:
\begin{Verbatim}[fontsize=\small]
> head(ltree1)
           y        x1        x2       x3 x4
1 -53.006195  5.751550 23.450707 2.058269  a
2 -47.699393 15.766103 24.700083 9.425390  c
3 -57.718613  8.179538 11.407989 3.793238  a
4 -69.926308 17.660348  5.765374 6.262401  a
5   7.747683 18.809346 17.387232 1.835024  a
6 -44.152011  0.911130 13.915809 6.592076  b
\end{Verbatim}
\noindent The fitted tree is obtained by

\begin{Verbatim}[fontsize=\small]
> fit_tree <- ltree(ltree1$y,
                    ltree1[, c("x1", "x2", "x3", "x4")],
                    trm = 0.1, max_depth = 10, prune = TRUE)
\end{Verbatim}
\noindent {In this call, \texttt{partition\_vars} is not supplied separately, so the second argument is used both to construct the local linear fits and to define the candidate split variables. The numeric variables enter the node-specific linear regressions, while the categorical variable \texttt{x4} is treated as a categorical partitioning variable and is evaluated through binary partitions of its levels.}
\noindent The fitted tree can also be visualized directly by

\begin{Verbatim}[fontsize=\small]
> plot(fit_tree)
\end{Verbatim}

\begin{figure}[h!]
\centering
\includegraphics[width=0.9\textwidth]{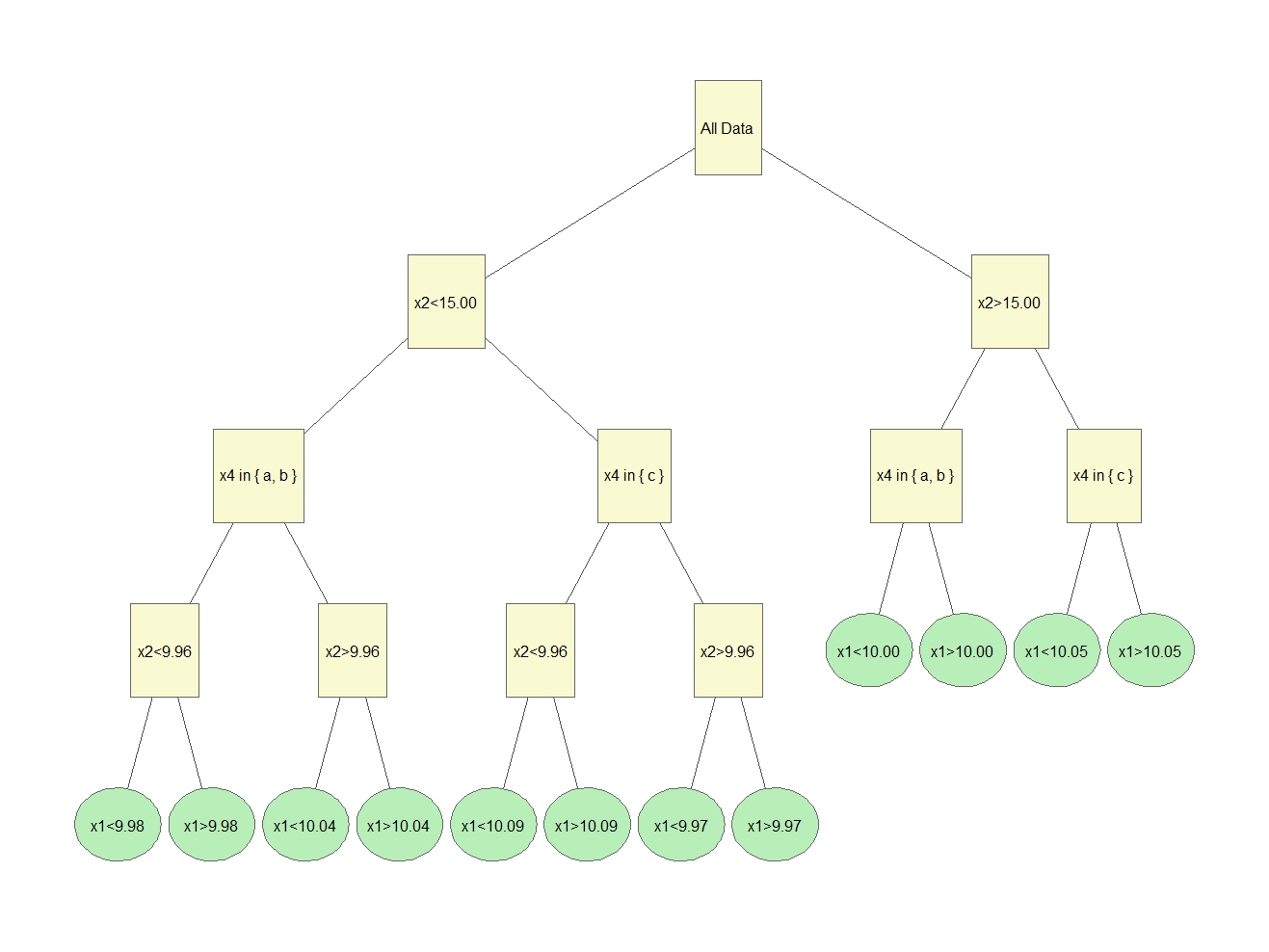}
\caption{Tree plot for Example 4.3.1.}
\label{fig:ltree-example1}
\end{figure}

\noindent Figure \ref{fig:ltree-example1} shows split locations close to the true thresholds, namely \texttt{X1 = 10}, \texttt{X2 = 10} and \texttt{15}, together with the categorical split \texttt{\{a,b\}} versus \texttt{\{c\}}. This example illustrates how the method can recover interpretable partition rules from a linear tree design.

\noindent \textbf{Example 4.3.2: smooth surface approximation.} Consider the smooth surface
\[
Y = m(X) + \epsilon, \quad \epsilon \sim N(0, 0.01), \quad X_1, X_2 \sim U(-1, 1)
\]
where,
\[
m(X) = \max \left\{ e^{-10X_{1}^{2}}, e^{-50X_{2}^{2}}, 1.25e^{-5(X_{1}^{2}+X_{2}^{2})} \right\}.
\]
One sample from this design is included in the package as \texttt{ltree2} and can be loaded using

\begin{Verbatim}[fontsize=\small]
> data(ltree2)
\end{Verbatim}

\noindent The data can be inspected by:
\begin{Verbatim}[fontsize=\small]
> head(ltree2)
           y         x1           x2
1 1.12789941  0.1767899 -0.008923763
2 0.01103560  0.8935761  0.318671184
3 -0.02411579  0.9185247 -0.949628182
4 0.80749526 -0.1300537 -0.346489341
5 0.89763768  0.1842471 -0.391250508
6 0.91290455  0.1418571 -0.522094958
\end{Verbatim}

\noindent The model is estimated from this sample by

\begin{Verbatim}[fontsize=\small]
> fit_surface <- ltree(ltree2$y,
                       cbind(1, ltree2$x1, ltree2$x2),
                       partition_vars = ltree2[, c("x1", "x2")],
                       trm = 0.1, max_depth = 10,
                       min_obs = 10, prune = TRUE)
\end{Verbatim}

\noindent {This example separates the two roles explicitly. The matrix \texttt{cbind(1, ltree2\$x1, ltree2\$x2)} defines the regressors in the leaf-specific linear models, while \texttt{partition\_vars = ltree2[, c("x1", "x2")]} defines the variables over which the tree searches for splits.}

\noindent To visualize the fitted surface, predictions can be computed on a grid and plotted by

\begin{Verbatim}[fontsize=\small]
> grid_x <- seq(-1, 1, length.out = 50)
> grid <- expand.grid(x1 = grid_x, x2 = grid_x)
> grid$z_ltree <- predict(fit_surface,
                          cbind(1, grid$x1, grid$x2),
                          new_partition_vars = grid[, c("x1", "x2")])
> grid$z_true <- pmax(exp(-10 * grid$x1^2),
                      exp(-50 * grid$x2^2),
                      1.25 * exp(-5 * (grid$x1^2 + grid$x2^2)))
> z_ltree_mat <- matrix(grid$z_ltree, nrow = 50)
> z_true_mat <- matrix(grid$z_true, nrow = 50)
> persp(grid_x, grid_x, z_true_mat, theta = 30, phi = 20,
  main = "continuous surface", xlab = "X1", ylab = "X2", zlab = "m(X)",
  col = "lightblue", shade = 0.3)
> persp(grid_x, grid_x, z_ltree_mat, theta = 30, phi = 20,
  main = "ltree approximation", xlab = "X1", ylab = "X2", zlab = "m(X)",
  col = "lightgreen", shade = 0.3)   
\end{Verbatim}

\begin{figure}[htbp]
\centering
\subfloat[Continuous surface]{
\includegraphics[width=0.47\textwidth]{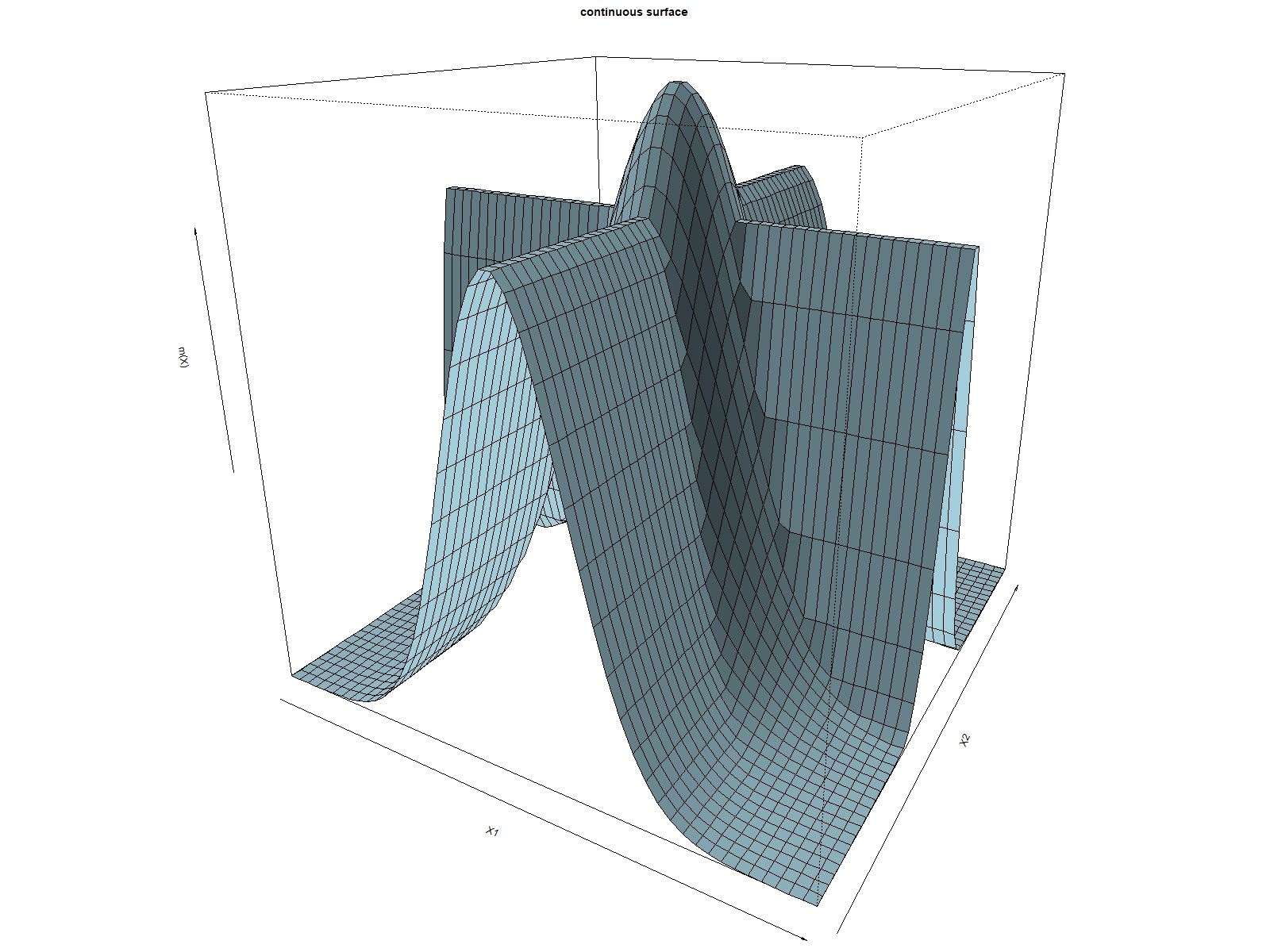}
}
\subfloat[Regression tree approximation]{
\includegraphics[width=0.47\textwidth]{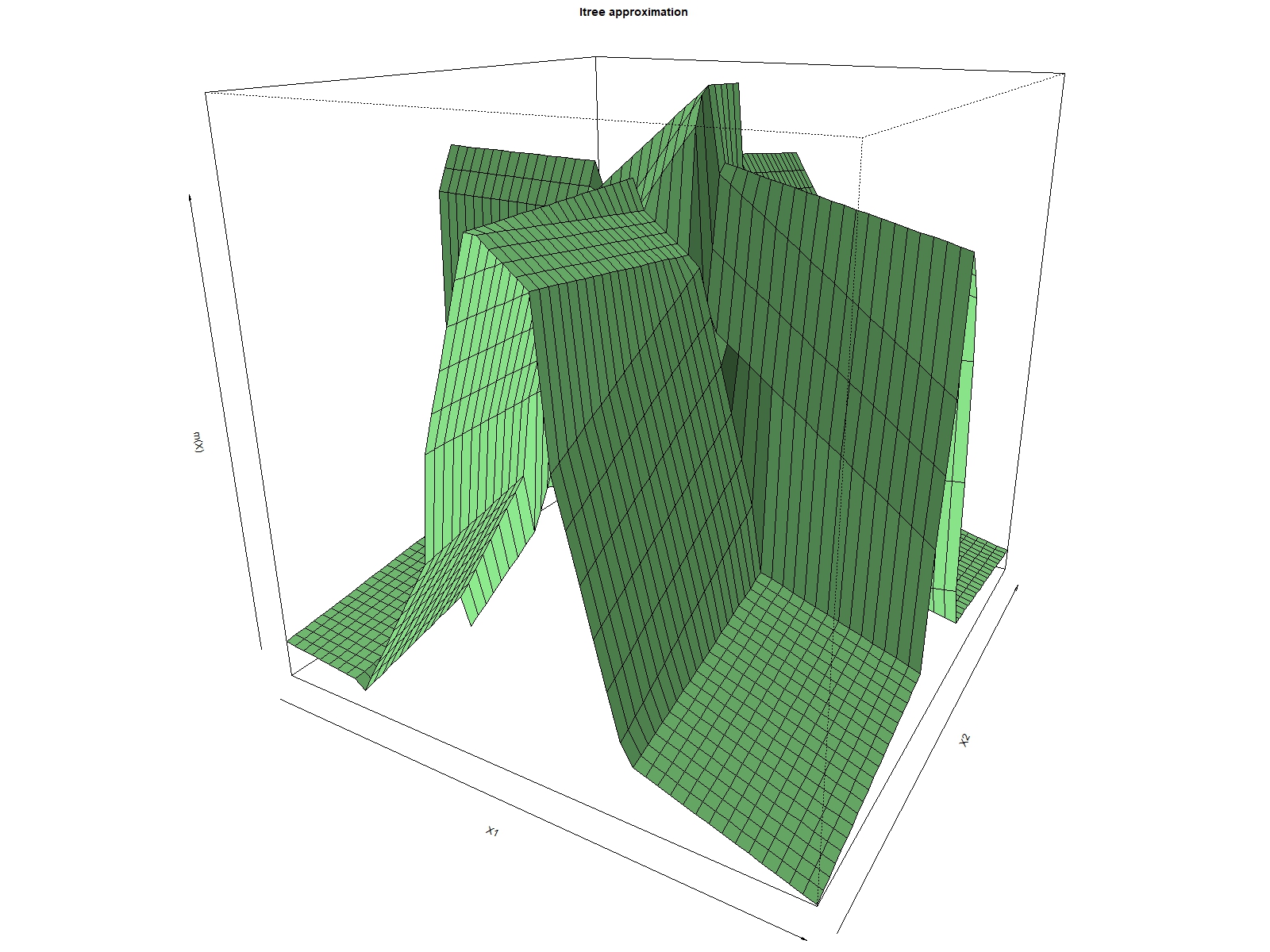}
}
\caption{Continuous surface and regression tree approximation for Example 4.3.2}
\label{fig:ltree-example2}
\end{figure}

\noindent Figure \ref{fig:ltree-example2} compares the continuous surface with the fitted surface implied by the tree approximation. This example illustrates how the generalized regression tree can approximate a smooth nonlinear regression function through local linear fits. Taken together, the two examples show that the procedure can be used both to recover interpretable partition rules and to approximate nonlinear regression functions.

\section{Conclusion}

The paper has introduced the \texttt{rbreak} package, which implements methods for detecting structural breaks and estimating break locations in linear multiple regression models under general linear restrictions on the coefficient vector. The package provides break date estimation with confidence intervals, a restricted sup-F test for the null of no structural change, simulation of critical values by Monte Carlo, and a bootstrap restart procedure to reduce the risk of convergence to spurious local optima. It also implements a generalized regression tree procedure, where each leaf contains a linear regression model rather than a local average. Future research can extend this framework beyond the single-equation setting to multiple-equation systems, which is currently in progress.

\newpage

\baselineskip=15.5pt

\bibliographystyle{agsm}
\bibliography{references}
\end{document}